\documentclass{pasa}%

\usepackage{color}

\newcommand{\putfig}[2]{\includegraphics[width=#2]{#1.pdf}}
\newcommand\degr{\hbox{$^\circ$}}
\newcommand\arcmin{\hbox{$^\prime$}}
\newcommand\arcsec{\mbox{$^{\prime\prime}$}}%
\newcommand\HI{\textsc{Hi}\ }

\newcommand\PKSB{PKS\,B1934--638}

\title[The Australian Square Kilometre Array Pathfinder]{The Australian Square Kilometre Array Pathfinder: Performance of the Boolardy Engineering Test Array}

\author[McConnell et al.]{D. McConnell$^{1}$, J.R. Allison$^{1}$, K. Bannister$^{1}$, M.E. Bell$^{1}$, H.E. Bignall$^{2,3}$, A.P. Chippendale$^{1}$, P.G. Edwards$^{1}$, L. Harvey-Smith$^{1}$, S. Hegarty$^{4}$, I. Heywood$^{1,5}$, A.W. Hotan$^{3}$, B.T. Indermuehle$^{1}$, E. Lenc$^{6,7}$, J. Marvil$^{1}$, A. Popping$^{8}$, W. Raja$^{1}$, J.E. Reynolds$^{1}$, R.J. Sault$^{1,9}$, P. Serra$^{1}$, M.A. Voronkov$^{1}$, M. Whiting$^{1}$, S.W. Amy$^{1}$, P. Axtens$^{1,10}$, L. Ball$^{1}$, T.J. Bateman$^{1}$, D.C.-J. Bock$^{1}$, R. Bolton$^{1}$, D. Brodrick$^{11,12}$, M. Brothers$^{1}$, A.J. Brown$^{1}$, J.D. Bunton$^{1}$, W. Cheng$^{1}$, T. Cornwell$^{1,13}$, D. DeBoer$^{1,14}$, I. Feain$^{1,15}$, R. Gough$^{1}$, N. Gupta$^{1,16}$, J.C. Guzman$^{3}$, G.A. Hampson$^{1}$, S. Hay$^{17}$, D.B. Hayman$^{1}$, S. Hoyle$^{18}$, B. Humphreys$^{1}$, C. Jacka$^{1}$, C.A. Jackson$^{1,2}$, S. Jackson$^{19}$, K. Jeganathan$^{1}$, J. Joseph$^{17}$, B.S. Koribalski$^{1}$, M. Leach$^{1}$, E..S. Lensson$^{1}$, A. MacLeod$^{1}$, S. Mackay$^{1}$, M. Marquarding$^{1}$, N.M. McClure-Griffiths$^{1,20}$, P. Mirtschin$^{12}$, D. Mitchell$^{1}$, S. Neuhold$^{1}$, A. Ng$^{1}$, R. Norris$^{1,21}$, S. Pearce$^{1}$, R.Y. Qiao$^{17,22}$, A.E.T. Schinckel$^{1}$, M. Shields$^{1}$, T.W. Shimwell$^{1,23}$, M. Storey$^{1}$, E. Troup$^{1}$, B. Turner$^{18,24}$, J. Tuthill$^{1}$, A. Tzioumis$^{1}$, R.M. Wark$^{1}$, T. Westmeier$^{1,8}$, C. Wilson$^{1}$, T. Wilson$^{12}$\\ 
\affil{$^{1}$CSIRO Astronomy and Space Science, PO Box 76, Epping NSW 1710, Australia}
\affil{$^{2}$International Centre for Radio Astronomy Research (ICRAR), Curtin University, Bentley WA 6102, Australia}
\affil{$^{3}$CSIRO Astronomy and Space Science, PO Box 1130, Bentley WA 6102, Australia}
\affil{$^{4}$Centre for Astrophysics \& Supercomputing, Swinburne University of Technology, PO Box 218, Hawthorn, Victoria, 3122, Australia }
\affil{$^{5}$Department of Physics and Electronics, Rhodes University, PO Box 94, Grahamstown, 6140, South Africa}
\affil{$^{6}$ARC Centre of Excellence for All-sky Astrophysics (CAASTRO)}
\affil{$^{7}$Sydney Institute for Astronomy, School of Physics, University of Sydney, NSW 2006, Australia}
\affil{$^{8}$International Centre for Radio Astronomy Research (ICRAR), University of Western Australia, Crawley, WA 6009, Australia}
\affil{$^{9}$School of Physics, University of Melbourne, VIC 3010, Australia}
\affil{$^{10}$Broadcast Support, TX Australia Pty Ltd, PO BOX 135, North Ryde BC NSW 1670, Australia}
\affil{$^{11}$ESS, Box 176, 221 00 Lund, Sweden}
\affil{$^{12}$CSIRO Astronomy and Space Science, 1828 Yarrie Lake Road, Narrabri NSW 2390, Australia}
\affil{$^{13}$Tim Cornwell Consulting, 9 Chapel Street, Sandbach CW11 1DS, United Kingdom}
\affil{$^{14}$Radio Astronomy Laboratory, University of California Berkeley, 501 Campbell, Berkeley CA 94720-3411, USA}
\affil{$^{15}$School of Medicine Radiation Physics Laboratory, University of Sydney NSW 2006, Australia}
\affil{$^{16}$Inter-University Centre for Astronomy and Astrophysics, Post Bag 4, Ganeshkhind, Pune University Campus, Pune 411 007, India}
\affil{$^{17}$CSIRO Digital Productivity, PO Box 76, Epping NSW 1710, Australia}
\affil{$^{18}$CSIRO Astronomy and Space Science, PO Box 276, Parkes NSW 2870, Australia}
\affil{$^{19}$CSIRO Astronomy and Space Science, PO Box 2102, Geraldton WA 6531, Australia}
\affil{$^{20}$Research School of Astronomy and Astrophysics, The Australian National University, Canberra, ACT 2611, Australia}
\affil{$^{21}$Western Sydney University, Locked Bag 1797, Penrith South, NSW 1797, Australia}
\affil{$^{22}$Sonartech ATLAS Pty Ltd,  Unit G01, 16 Giffnock Avenue, Macquarie Park NSW 2113, Australia}
\affil{$^{23}$Leiden Observatory, Leiden University, PO Box 9513, NL-2300 RA  Leiden, The Netherlands}
\affil{$^{24}$Charles Sturt University, Locked Bag 588, Wagga Wagga NSW 2678, Australia}
}

\jid{PASA}
\doi{10.1017/pas.\the\year.Xx}
\jyear{\the\year}

\usepackage[authoryear]{natbib}
\bibpunct{(}{)}{;}{a}{}{,}
\begin{document}%

\begin{abstract}

We describe the performance of the Boolardy Engineering Test Array (BETA), the
prototype for the Australian Square Kilometre Array Pathfinder telescope
ASKAP. BETA is the first aperture synthesis radio telescope to use phased
array feed technology, giving it the ability to electronically form up to nine
dual-polarization beams. We report the methods developed for forming and
measuring the beams, and the adaptations that have been made to the
traditional calibration and imaging procedures in order to allow BETA to
function as a multi-beam aperture synthesis telescope. We describe the
commissioning of the instrument and present details of BETA's performance:
sensitivity, beam characteristics, polarimetric properties and image quality.
We summarise the astronomical science that it has produced and draw lessons
from operating BETA that will be relevant to the commissioning and operation
of the final ASKAP telescope.

\end{abstract}


\maketitle%


\section{INTRODUCTION}
\label{sec:intro}

For more than two decades, the astronomical community has recognised the need for a major new radio
observatory to succeed the current generation of radio telescopes, and to surpass their sensitivity
and resolution. The Square Kilometre Array (SKA) was conceived in the early 1990s
\citep{Carilli:2004ww} and is now embodied in The SKA Organisation\footnote{www.skatelescope.org},
which is coordinating the design and construction of major new radio telescopes.

ASKAP, the Australian SKA Pathfinder (\citeauthor{DeBoer:hj}
\citeyear{DeBoer:hj}, \citeauthor{Johnston:2007ku}
\citeyear{Johnston:2007ku}), is one of several radio telescopes being built to
explore and demonstrate possible new approaches to designing the SKA itself.
ASKAP is located at the Murchison Radio Observatory (MRO) in Western
Australia, and is being constructed by CSIRO---the Commonwealth Scientific and
Industrial Research Organisation. It is designed as a survey telescope that
can rapidly image the entire available sky. It operates over the 0.7 --
1.8~GHz range and will achieve its survey speed by virtue of small antennas
(12~m diameter) and the use of phased array feeds (PAFs) to sample a large
portion (approximately 0.64 m$^2$) of the focal plane. Other SKA precursors
include the Murchison Widefield Array (MWA) \citep{Tingay:2013ef}, also
located at the MRO, and the South African array MeerKAT \citep{Jonas:2009gc}.

The phased array feed is the critical new technology being explored on ASKAP.
A PAF is a dense array of sensors placed in the focal plane of each antenna.
Digital beamformers synthesize `formed' beams as linear
combinations of signals from the individual sensors. With a suitable choice
of weights---the complex coefficients used in the linear sum---beams can be
formed to point anywhere within the available field-of-view, 5.5\degr $\times$
5.5\degr\ in the case of ASKAP. \cite{Verheijen:2008aa} describe APERTIF
(APERture Tile In Focus), another exploratory project, which is equipping
antennas of the WSRT (Westerbork Synthesis Radio Telescope) with PAFs.

Here we describe the operation and performance of a prototype of ASKAP, the
Boolardy Engineering Test Array (BETA), comprising six of the ASKAP antennas.
The full description of BETA by \citet{Hotan:2014dv} is recommended reading to
set the context of this account of its performance.

The reliance of ASKAP on such a novel technology has risks, many of which may
be reduced by the establishment of BETA as a functioning prototype capable of
astronomical observations. Questions to be answered by experiment with BETA
include:

\begin{itemize}

\item Can beams be formed with only small antennas? The beams are formed as a linear combination
of the signals from individual PAF sensors, specified as a set of complex
weights. Determination of the optimum set of weights requires a measure of the response of each
sensor, which in turn requires a sufficiently strong source of radiation to provide a measurable
response.

\item Are the beams stable, and are the time scales for beam degradation long compared with the practical interval for
determination of beam weights?

\item Can the antenna pointing be measured? Traditionally, the single stable beam aligned with the antenna's optical axis provides
the reference for the assesment of mechanical pointing accuracy.

\item Can variations in the antenna-to-antenna beam shape be made small enough not to limit dynamic range?
\item Can the instrumental polarization be calibrated and corrected over the whole field of view?
\item Is the expected field-of-view realised?
\item Can the antenna gains (beam and frequency specific) be determined with a small number of small antennas?
\item Can an instrument of this complexity operate with the reliability required for uniform all-sky surveys?
\end{itemize}

\paragraph{Outline}

After giving a brief description of BETA and its commissioning
(\textsection\ref{BETA}), we follow a logical sequence through the methods and
results of its operation: calibration of the array (antenna locations and
pointing) in \textsection{\ref{instrumental}}; beamforming and beam
measurement in \textsection{\ref{beamforming}}; observations, including those
required for calibration of antenna gains in \textsection{\ref{observations}};
calibration of visibility data for the instrument's response and imaging those
data in \textsection{\ref{calibration}}. In \textsection{\ref{performance}} we
present details of BETA's perfomance: sensitivity, beam characteristics,
polarimetric properties and image quality. In \textsection{\ref{summary}} we
briefly summarise the use of BETA and point to the future operation of ASKAP,
drawing on the experience gained with BETA.

%

\section{COMMISSIONING AND FIRST SCIENCE}
\label{BETA}

\subsection{BETA the telescope}

BETA operated as a aperture synthesis telescope from March 2014 until it was
decommissioned in February 2016. The parameters of BETA are summarised in
Table \ref{BETAparameters}, reproduced from \citet{Hotan:2014dv}. The BETA PAF
is a dual-polarization connected-array antenna (\citeauthor{Hay:2008fb}
\citeyear{Hay:2008fb}, \citeauthor{Hay:hp} \citeyear{Hay:hp}) with 2$\times$94
sensors in a `chequerboard' pattern. It is the first version (Mark I) of the
CSIRO PAF \citep{Schinckel:2011}, and will be superceded in the final
ASKAP telescope by the Mark II, which has better and more consistent noise
properties across its frequency range \citep{Chippendale:2015wyex}.

BETA comprises six of the 36 ASKAP antennas. Each has conventional vertical
and horizontal (azimuth and elevation) rotation axes, and a third axis---the
`roll' axis---that is colinear with the antenna's optical axis or `boresight'. Motion
about the roll axis allows the antenna's response pattern to be kept fixed on the
sidereal sky.

For each antenna, the digital beamformers can synthesize nine
dual-polarization beams at each of the 304 coarse (1-MHz) frequency channels across
the observed band. After further division of the spectrum into 16416 fine
frequency channels, the correlator computes, for all four polarisation
products (XX, XY, YX, YY), visibilities for all nine beams across the
6-antenna array.

BETA implements phase and delay tracking for a reference direction, common to
all beams. This approach is feasible for BETA since all baselines are shorter
than 1000~m. ASKAP will phase-track a reference direction for each beam.

BETA lacks a system for injecting noise into the radio frequency signal. Such
a system, which will be present on ASKAP, would be used for stabilising gain
variations and for measuring the relative phase of signals in the X and Y
polarization channels. BETA has been operated without any gain stabilisation,
and for polarimetric work the XY phase has been estimated by purposely
misaligning one antenna about its roll axis.

A major component of ASKAP is ASKAPsoft \citep{Cornwell:2011}, the calibration
and imaging software package that processes raw visibility data to produce
images of the sky. Although the number of antennas and sensitivity of BETA was
insufficient to test the full sky-model based calibration approach designed
for ASKAP, it provided a means to test the software with more traditional
calibration methods, and has allowed direct comparison with existing radio
astronomy imaging packages.

The use of phased array feeds greatly increases hardware complexity compared
to other radio telescopes. With 188 active elements at the focus, the task of
monitoring and control becomes a significant engineering exercise, the more so
for needing to operate the telescope remotely. BETA has provided a platform
to test the systems that were designed to cope with this complexity. Issues of
scalability, logging efficiency and the provision of adequate user feedback
all arose during the operation of BETA and have influenced the design of the
final system to be used on ASKAP.

\begin{table}[h]
\begin{center}
\caption[]{Key parameters of the BETA telescope.}
\begin{tabular}{ll}
 \hline
 \hline
Number of antennas & 6 \\
Antenna diameter & 12 m \\
Total collecting area & 678.6 m$^2$ \\
Maximum baseline & 916 m\\
Angular resolution & 1.3 arcmin\textdagger \\
Observing frequency & 0.7--1.8 GHz\\
Simultaneous bandwidth & 304\,MHz \\
Frequency channels & 16416\\
Frequency resolution & 18.5 kHz\\
Simultaneous beams & 9 (dual-pol) \\
Minimum integration time & 5 s\\
\hline
\hline
\end{tabular}\\
\textdagger {\footnotesize Natural weighting, 1.1-1.4~GHz } \\
\label{BETAparameters}
\end{center}
\end{table}

\subsection{Operational performance}

ASKAP is built on a remote and isolated site. The Murchison Radio Observatory
is located in a sparsely populated region of Western Australia, 380~km north-east 
by road from the town of Geraldton. Support staff are based in
Geraldton and, while construction continues, travel to the Observatory by road
or air for several days of most weeks. The Observatory is unstaffed at other
times, and so the establishment of reliable operations is essential to the
efficiency of ASKAP's scientific program. Most operations can be conducted
remotely.

BETA has proven to be workable in this mode of operation, but several
reliability issues highlighted the importance of developing robust systems for
deployment in this harsh environment. The most serious examples were the MRO
site power supply, and certain components of the systems for cooling PAFs and
other antenna-based equipment.

Observatory power is provided by a set of diesel motor-generators, designed to
be swapped without interruption to power. Over the commissioning period the
reliability of the power has steadily improved; interruptions experienced in
2015 were 10-fold fewer than in 2014.

The antenna-based electronics was cooled by air circulated through a water-cooled
heat exchanger. Variation of water flow rate was used to stabilise the
temperature of the equipment enclosures. These cooling systems, which usually
ran at close to their capacity, gave a range of difficulties: water
circulation pump burn-out; false alarms from over-temperature sensors; and
power irregularities induced by start-up transients from the compressors.
Although the mean time between failures in either site power or PAF cooling
has been moderate (approximately 30 days), the impact on operations is greater
than for failures in most other parts of the telescope because of the need for
support staff to restore normal operation. Cooling systems for the Mark II
PAFs and associated equipment have been redesigned.

The telescope performance has also proven to be relatively stable. We present results in
\textsection\ref{performance} that are evidence for a pleasing level of
stability in antenna pointing, in amplitude, phase and delay of the signal
path, and in instrumental polarization. The most significant areas of
instability are an occasional loss of synchronisation in the correlator (eg.
\citeauthor{2015MNRAS.453.1249A} \citeyear{2015MNRAS.453.1249A}), a diurnal
gain-amplitude variation arising from imperfect PAF temperature stabilisation
and some loss of coherence between PAF digital sample streams. The last issue
is discussed in \textsection\ref{results_beams}

\subsection{Interfering signals}

ASKAP and the MRO are protected from ground-based sources of radio frequency
interference (RFI) by the Australian Radio Quiet Zone WA (ARQZWA), established
by the Australian and Western Australian Governments
\citep{Wilson:2013,ACMA:2011}.  The Zone has three concentric parts with radii
of 70km, 150km and 260km, which have graded regulations designed to minimise
radio interference from ground-based sources.  As a consequence of this
protection and of the very low population density throughout the Zone, most
RFI detected by BETA originated from navigation and communications satellites
and from aircraft navigation systems.

The BETA spectrum shown in Figure \ref{sefd} has several features identified
with these airborne and orbiting sources of RFI.  Allison et al. (in prep)
show the spectrum in more detail: spectra over the 711.5--1527.5 MHz band from
two astronomical sources have very similar patterns of RFI that are consistent with
the spectrum in Figure \ref{sefd}. They report that about 14 per
cent of both spectra are corrupted by RFI.

\subsubsection{Solar interference}

\begin{figure}[h]
\begin{center}
\putfig{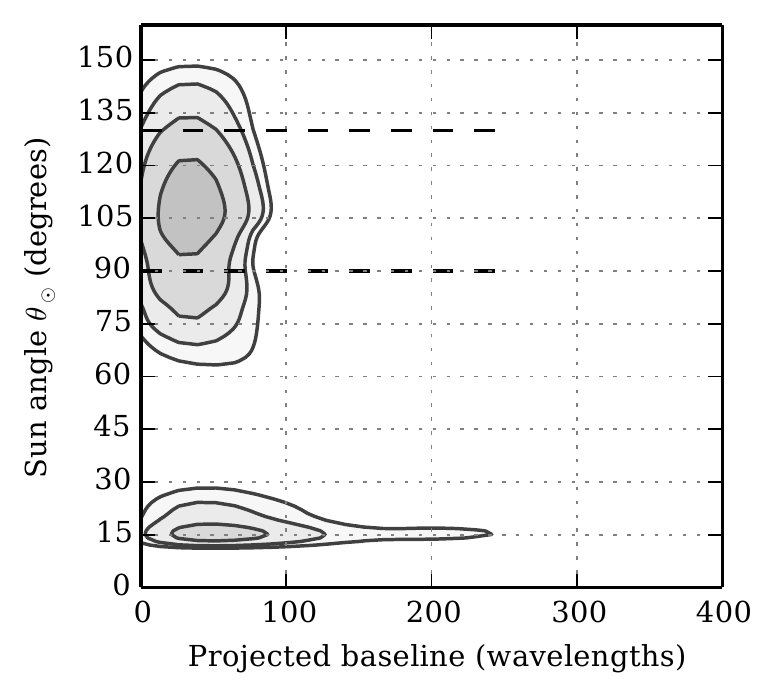}{\columnwidth}
\caption{
The strength of solar interference detected in BETA visibilities as a function
of projected baseline length and the Sun's displacement $\theta_{\odot}$ from
the pointing direction. When $\theta_{\odot}$ lies between the two dashed
lines, the Sun shines directly onto the surface of the PAF. At greater angles,
the PAF is shadowed by the antenna's primary reflector. The contours are at 5, 10, 20 and 40 Jy.}

\label{solar}
\end{center}
\end{figure}

The Sun is a source of radio interference and appears in BETA visibilities as
broad-band noise with phase structure characteristic of a source displaced
from the Array's delay tracking centre \citep{Hotan:2014dv}.  Over the 711.5--
1015.5~MHz band, we explored the circumstances that led to the greatest levels
of solar interference by recording visibilities from directions over a
coarsely-spaced ($\sim15\degr$) range of angular displacements
$\theta_{\odot}$ from the Sun. The spectrum of each visibility measurement was
Fourier-transformed to produce a delay spectrum in which the solar signal is
easily identified at the delay corresponding to its angular displacement from
the delay centre.  We estimated the absolute strength of the solar signal by
comparison with a contemporaneous observation of the source PKS B0407--658,
which has a flux density of 24.4Jy at 843~MHz \citep{Mauch:2003ed}. The
results are summarised in Figure \ref{solar}.  The solar signal is greatest on
short baselines; the Sun's 0.5\degr\ disk is resolved for baselines much
longer than $\sim60\lambda$. Solar interference is greatest when
$\theta_{\odot} \lesssim 20\degr$ and $80\degr \lesssim \theta_{\odot}
\lesssim 130\degr$. When $\theta_{\odot}$ lies in the range
[90\degr,130\degr], the sun shines directly onto the surface of the PAF; at
greater angles the PAF is shadowed by the antenna's primary reflector.

\subsection{First science}
\label{firstscience}

The scientific program of ASKAP is based on a set of Survey Science Projects
(SSP) that cover a broad range of radio surveys of the sky
\citep{Johnston:2007ku}. Up to 75 per cent of ASKAP time will be dedicated to these
projects. Although BETA, with its few 12-m antennas and the relatively poor
sensitivity of the Mark I PAFs at frequencies $\gtrsim$1.2~GHz, was not
expected to produce major scientific contributions, the wide instantaneous
field of view and access to the 0.7--1.8~GHz portion of the spectrum from a
radio-quiet site gave the potential for some scientifically useful results.
Therefore the commissioning team included active members of several of the
SSPs, adding scientific incentive to the motivation for the commissioning
activities and resulting in six refereed scientific publications.

The higher-sensitivity part of BETA's spectrum corresponds to frequencies of
the \HI line at redshifts of $0.4 < z < 1.0$. \citet{2015MNRAS.453.1249A}
report the first detection of strong \HI absorption at $z = 0.44$ associated with the
young radio galaxy PKS1740-517. Contrary to the normal sequence, the redshift
of this system was measured first in the radio, and later confirmed in optical
spectra.

\citet{Serra:2015gz} report observations at 1.4 GHz of the galaxy group IC 1459. BETA's
nine beams at 1.4GHz covered 6 square degrees, within which \HI emission was detected from 11 of the
galaxies in the group. Three previously undetected \HI clouds were discovered in the BETA images.

One of the scientific objectives of ASKAP is to detect transient radio emission on various time
scales. To explore this capability, BETA was used to image a field containing the intermittent pulsar PSR
J1107-5901 already known to exhibit marked `on' and `off' states. The result was the
successful detection of several state transitions in a series of 390 2-minute images, reported
in full by \cite{Hobbs:2016hi}.

Much of BETA's commissioning work has focussed on continuum imaging of large areas of sky, a
capability required by several of ASKAP's SSPs. \citet{Heywood:2016di} present
the results from repeated observations of a 150-square degree field in the constellation Tucana.
Three 12-hour observations were made, with each observation being two sets of six
pointings, offset to achieve the interleaving described below in \textsection{\ref{observations}}. From
the resulting images, \citeauthor{Heywood:2016di} prepared a new catalogue of 3,722 sources, giving
positions, flux densities and, in many cases, spectral indices over the 711.5--1015.5\,MHz observing band.

\cite{Harvey-Smith:2016aa} report detection by BETA of emission from the OH
megamaser IRAS~20100-4156, confirmed by subsequent observations with the ATCA
(Australia Telescope Compact Array). Their paper discusses the significance
of changes in the emission spectrum since the first observation of this source
by \cite{1989Natur.337..625S}.

BETA was used for part of the comprehensive search for an
electro-magnetic counterpart to the gravitational wave (GW) event GW150914
reported by \cite{Abbott:2016cn}.  BETA's contribution to this campaign
\citep{Abbott:2016wm} was the imaging of approximately 270 square degrees of the
most likely GW localization. Observations were made with BETA over 18 hours about one
week after the GW event, and reached an image sensitivity of 1-5 mJy (rms).

\section{INSTRUMENTAL CALIBRATION}
\label{instrumental}

\subsection{Antenna positions}

All ASKAP antenna locations in the horizontal plane were initially determined
by conventional surveying techniques, with the vertical distance from the WGS
84 reference spheroid\footnote{World Geodetic System 1984} also measured for
the six BETA antennas. A standard method for refining these measurements is to
observe several strong radio sources with well-known celestial positions, and
use the interferometer phase on each baseline to derive corrections to the
initial measurements. We used an equivalent and, for BETA, more robust
technique of performing self-calibration of the visibilities for each source
and fitting each antenna's position ($X, Y, Z$ in the International
Terrestrial Reference Frame) to the phase of its complex gain. We used a
simple point-source model of the field, but more complex models could
have been used to allow calibration on confused fields. The observations were
made using single-element beams to avoid measurement bias introduced by the
comparatively poorly-understood complex gains of the beams formed from many
elements. The uncertainties in position corrections determined in this way are
typically 0.5~mm in the $X-Y$ plane, and 1.3~mm in $Z$.

Once the antenna locations were established, fixed (or slowly varying)
antenna-specific delays were determined from a single observation of a strong
source. These delays are determined with an accuracy of $\sim$50~ps.

\subsection{Antenna pointing}
\label{antennapointing}

In a radiotelescope with a feed horn, the beam direction is fixed relative to the telescope's
structure and defines its pointing direction. Imperfections in the mechanical pointing can be
determined by measuring the beam position on the sky relative to the known position of a
strong source---a pointing calibrator. Any misalignment between the beam and the telescope's
optical axis is ultimately absorbed in the pointing model.

The ASKAP antennas form beams from a PAF; their direction relative to the
antenna's optical axis, and their shape, are determined by the weights used in
the combination of the signal from each PAF element and by the (possibly time-variable) element
characteristics. This introduces new sources of pointing error not related to
the mechanical pointing performance of the antenna.

Two methods have been developed for pointing measurements on BETA, both using unique features of the
ASKAP antennas. The first, used to determine coarse offsets on the azimuth and elevation motions,
determines the location of the Sun's image on the focal plane from the PAF element outputs. This
method uses the antenna and its PAF as a radio `digital camera'. The image quality on the focal plane
is imperfect because of the dispersion of PAF element characteristics, but is sufficient to correct
the coarse pointing to an accuracy of $\lesssim 0.1 \degr$.

The second method uses the roll rotation axis of the ASKAP antennas as the
reference direction for pointing calibration. The procedure uses beams formed
from single PAF elements, which have fixed offsets from the boresight
direction, but individually have less sensitivity than formed beams. By
rotating the antenna being measured about its roll axis while pointing at a
calibration source, and with knowledge of the size of single-element beams,
the amplitudes of visibilities with a second fixed reference antenna can be
modelled to give the misalignment of the roll-axis with the source direction
---the pointing error. The procedure is executed twice for each calibration
source; half the antennas are used as reference for measuring the other half,
then these roles are reversed. Eight of BETA's nine beamformers are used to make
the measurements simultaneously with a symmetrical pattern of eight single
elements in both polarizations.
\begin{figure}[h]
\begin{center}
\putfig{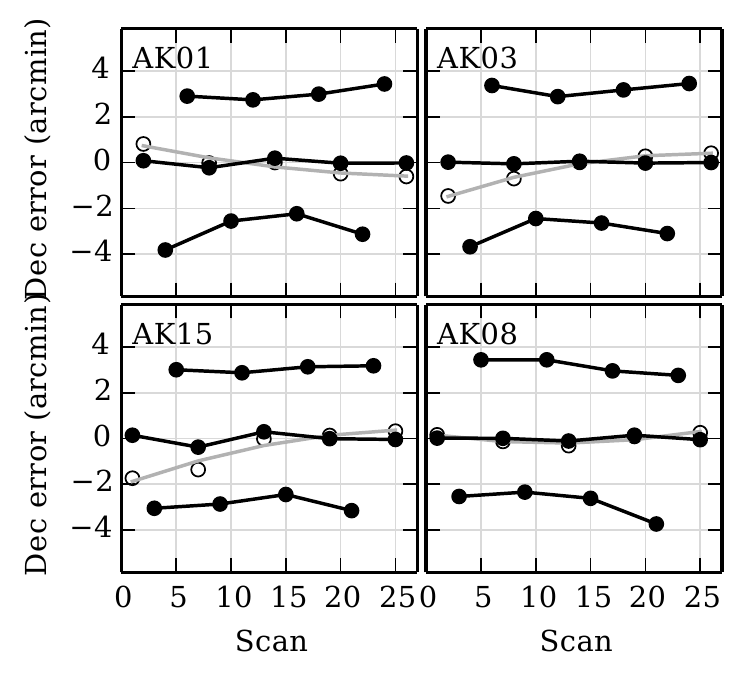}{\columnwidth}
\caption{
Results from the declination offset test that is described here, for antennas 1, 3, 8
and 15. The open grey points show the declination error measured with zero-offset
pointings. The grey line is the quadratic fit to these; the trend is
assumed due to an imperfect set of pointing model parameters. The filled black
symbols show the results with these trends removed, connected in groups for
the 0, $-3$ and $+3$ arcminute offsets. The abscissae number the 26 pointing
scans that extended over a 7-hour period.}
\label{pointingtest}
\end{center}
\end{figure}

Figure \ref{pointingtest} shows the results of observations made to test the
efficacy of this method. Pointing measurements were made repeatedly on a
single source (\PKSB) over a 7 hour period, with pointing intentionally set
with declination offsets of $-$3, 0 +3 arcminutes. Four antennas were used
for the test, with each alternately being measured or serving as reference. Twenty six
pointing scans were conducted over the 7 hours, resulting in 13 measurements
for each antenna. From these results we estimate the uncertainties in pointing
error determination to be $\Delta\texttt{HA}/\cos\delta = 0.4$, $\Delta\delta
= 0.3$ arcminutes. A full description of both methods can be found in the report by
\cite{McConnell:aces9}.

\subsubsection{Antenna roll axis}

An error in the setting of an antenna's roll axis angle results in pointing
errors for beams that are offset from the optical axis and introduces
polarization leakage in visibilities formed from all beams. Zero-point errors
in roll angle have been estimated using two different methods.  As for the
measurement of coarse antenna pointing described above, the location of the
Sun's image on the PAF was used. The antennas were held fixed on the meridian
at the Sun's declination, and analysis of the path followed by the Sun's image
across the PAF allowed the roll angle error to be measured. The second method
used observations designed to measure polarization leakage (see section
\ref{results_polarimetry} for more detail). Rotational misalignment of the PAF
produces symmetrical deviations in the real part of the leakage in the X and Y
polarized beams. Such deviations were observed and were used to estimate relative
roll angle errors.  The two methods gave consistent results; together they
resulted in roll angle accuracy of about 0.2\degr.


\section{BEAMFORMING AND MEASUREMENT}
\label{beamforming}

\subsection{The maxSNR beamforming method}

The beamforming practice developed and used with BETA follows closely the
methods described by \citet[section 5.2]{Hotan:2014dv} for forming beams that
maximise the signal-to-noise ratio (maxSNR) \citep{Applebaum:1976uf} in the
direction $\Omega$ chosen for the beam. The beamforming process is the
determination of the set of complex weights $\mathbf{w}_k$ for each beamformer
$k$, where the output of that beamformer at time sample $i$ is
\begin{eqnarray}
\mathbf{y}_k[i] & = & \mathbf{w}^H_k \mathbf{x}[i]
\end{eqnarray}
and $\mathbf{x}[i]$ is the vector of complex PAF element voltages for a single
frequency channel.  There are $2 n_\Omega \times n_f$ beamformers where
$n_\Omega$ is the number of beam directions (9 for BETA), and $n_f$ is the
number of frequency channels; the factor 2 allows for both polarizations. Thus
$k \in \{0, 1, \ldots, 2 n_\Omega n_f-1\}$. The array covariance matrix (ACM)
$\hat{\mathbf{R}}$ is computed by the PAF ACM correlator \citep{Hotan:2014dv}
as
\begin{equation}
\hat{\mathbf{R}}  = <\mathbf{x} \mathbf{x}^H>
\end{equation}
and is used to estimate the array response $\hat{\mathbf{v}}_k$ for a unit
magnitude plane wave incident from the direction of the $k^{\text{th}}$ beam.
ACMs are computed while pointing, in turn, at `empty' sky and at a strong
source offset to the desired beam direction to obtain $\hat{\mathbf{R}}_n$ and
$\hat{\mathbf{R}}_{n+s}$ respectively.  The array response
$\hat{\mathbf{v}}_k$ for a unit magnitude plane wave incident from the
direction of the desired beam, is estimated as the  dominant solution
$\mathbf{v}_{1}$ to the eigenvalue equation \citep{2008ISTSP...2..635J}
\begin{eqnarray}
\label{eigen}
\hat{(\mathbf{R}}_{n+s} - \hat{\mathbf{R}}_{n}) \hat{\mathbf{v}} & = & \lambda  \hat{\mathbf{v}} 
\end{eqnarray}
from which the maxSNR weights are computed as 
\begin{eqnarray}
\label{wgts}
\mathbf{w}_k & = & \hat{\mathbf{R}}_n^{-1} {\mathbf{v}}_{1k}
\end{eqnarray}
Finally, for each beam pointing $\Omega$ the weight vectors for all frequency
channels are adjusted to ensure a smooth variation of phase over the band:
\begin{eqnarray}
\label{phi}
\phi & = & \arg \mathbf{w}^H \cdotp \mathbf{w}_r \\
\label{wprime}
\mathbf{w}^\prime & = & e^{-i\phi} \mathbf{w}
\end{eqnarray}
where $\mathbf{w}_r$ is the weight vector for some reference channel $r$.

To summarise, the maxSNR beamforming procedure with BETA was:
\begin{enumerate}
\item specify a set of nine beam directions;

\item for each beam, point the antenna so that a strong source (the Sun or
Taurus-A) lies at the corresponding offset and record the ACM
$\hat{\mathbf{R}}_{n+s}$ for all frequencies, with an effective integration of
1.25s over a 1-MHz bandwidth;

\item point all antennas at a reference field, typically 15\degr\ south of the
Sun (or Taurus-A) to obtain $\hat{\mathbf{R}}_n$; receiver noise dominates the
PAF output for all but a small number of fields);

\item the array response vector $\mathbf{v}$ and hence the weight vector
$\mathbf{w}$ are determined for each beam and frequency from  (\ref{eigen})
and (\ref{wgts}) above;

\item the set of weights for each frequency channel are adjusted to give a smooth
phase variation across the frequency band using (\ref{phi}),(\ref{wprime});

\item the weights $\mathbf{w}^\prime$ are loaded into the digital beamformers.
\end{enumerate}

Although this method determines weights for all antennas simultaneously, the
antennas function independently; no use is made of the ability to form
interferometers between pairs of antennas. Some experiments using
interferometers to characterise PAF elements have been conducted, but are
expected to be more successful with the full ASKAP capabilities. The greater
number of antennas and beamformers will improve both the sensitivity and the
efficiency of measurement, making the interferometric approach more practical.

Beams were formed for each of the 304 coarse frequency channels; there are 304
beamformers for each of the 2-polarization$\times9$ beams. However, BETA
supported ACM download for only 64 coarse (1-MHz) frequency channels at a time,
which we distributed across the 304-MHz instantaneous band. The result was 64
sets of beam weights, each applied to a contiguous band of 4 or 5 coarse
frequency channels. A consequence of this was detectable discontinuities in
the beam shape variation and bandpass shape at edges of these contiguous bands
\citep{2015MNRAS.453.1249A}.

\subsection{Beam measurement procedures} With the flexibility to form
different types of beam by varying the complex weights applied to each PAF
element, it was important to establish a measurement technique that could
quantify beam characteristics. Having quantitative measures of beam
characteristics is a prerequisite for future optimisation to meet scientific
goals.

In order to fully characterise formed beams, we developed a holography
procedure that uses interferometric observations between one or more reference
antennas and a set of antennas under test---the target antennas. The reference
antenna is set to track a reference source---a bright point-like astronomical
source, typically Virgo A---for the duration of the test, and is loaded with
nine identical copies of a beam pointing in the boresight direction. This
allows for simultaneous holography of 9 dual-polarization beams on the target
antennas. The complex visibility on the reference--target baseline is a measure
of the beam in the direction of target pointing.

We define a square grid of points centred on the reference source and aligned
with the elevation and cross-elevation directions. A measurement is made at each
point of the grid. The grid spacing is roughly half the width of the beam at
the frequency in question to provide enough spatial resolution to determine
the beam's response to the reference source. The roll axis is kept fixed at
an angle of zero, so that we measure the beam shape with respect to the
antenna structure.

For the lowest frequency BETA band, 711.5--1015.5\,MHz, we typically used a
$15\times15$ grid of 225 points spaced by 0.6\degr\ to cover an 8$\times$8
degree square. Once the visibilities have been gridded, bivariate spline
interpolation is used to smooth the image prior to visualisation.  The
resulting two-dimensional beam pattern can be analysed to determine properties
such as width at half power, ellipticity and relative side lobe levels.
Results of beam measurements are presented (Fig. \ref{SB1772_Mosaic}) and
discussed in \textsection{\ref{results_beams}}.

\section{OBSERVING WITH BETA}
\label{observations}

Observations with a radio synthesis-imaging telescope are made by pointing all
its antennas at the field to be imaged and measuring the correlation between
signals from all pairs of antennas, yielding the complex visibility of the
field for each baseline. Occasional observations of sources with known flux density
and celestial position allow the complex gains of each antenna to be
determined and subsequent calibration of the visibilities. BETA, with its set
of formed beams, requires some variations to this basic scheme.

\begin{figure}[h]
\begin{center}
\putfig{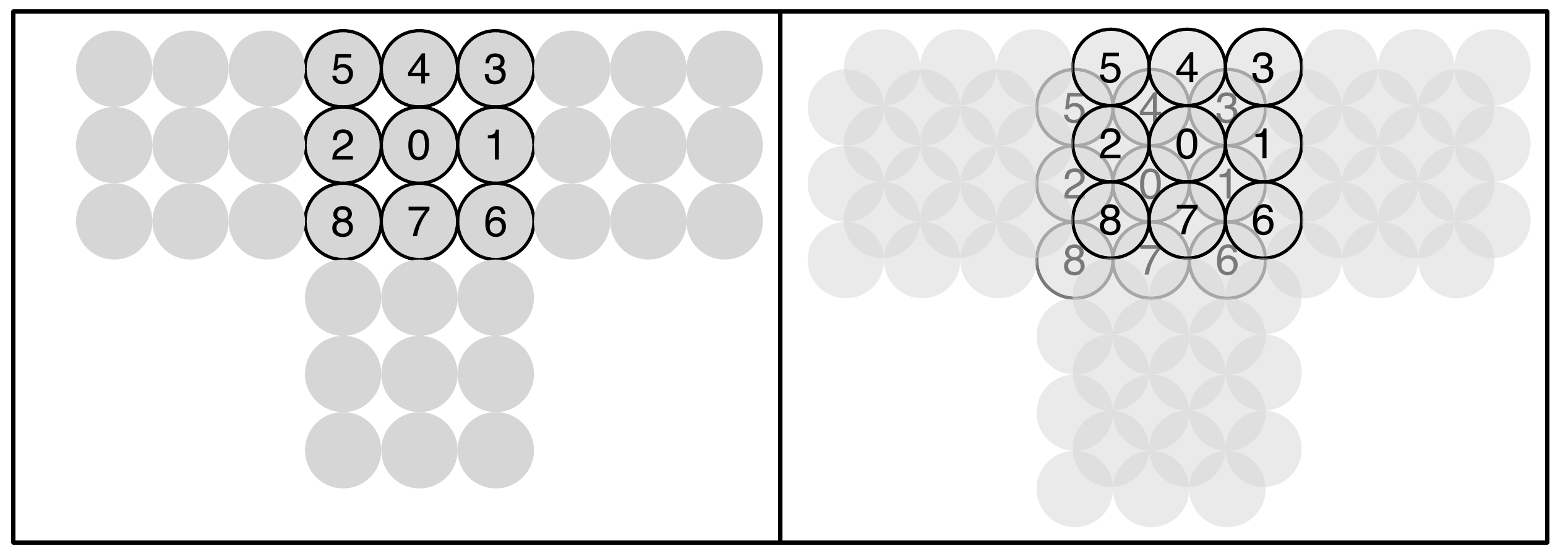}{\columnwidth}
\caption{
Left: a typical footprint used for BETA observations showing the locations of
beams 1--8 relative to the boresight beam 0 in the centre of the pattern; the
additional sky coverage shown is achieved by adjusting the antenna pointing
positions; four positions are used in this example. These are the `A'
pointings described in the text. Right: the same pattern, but with additional
interleaved `B' pointings. Each footprint is described by the name of its
geometry (in this case `square'), and its pitch, the spacing between beam
centres. A typical value for the pitch used for the 711.5--1015.5\,MHz band is
1.46 degrees, the approximate full-width at half maximum (FWHM) of the ASKAP
beam at the highest frequency in that band.}

\label{footprint}
\end{center}
\end{figure}

\paragraph{Beam footprints} The BETA PAFs and beamformers allow up to
nine beams to be formed to point anywhere within the $\sim$30 square
degree PAF field of view. The performance of each beam is determined by the
weights used and the characteristics of the PAF elements selected by those
weights. Therefore, it is important to make observations of both target and
calibration fields with the same set of beams. We define sets of beams by
their footprint, the arrangement of beam centres expressed as offsets from
the antenna optical axis. For a given footprint, beamformer weights are
determined (`beamforming'). These weights are then used for observations of
a flux-density calibrator with each beam in turn, and to measure the
visibilities of the field to be imaged. Footprints are designed to satisfy the
requirements of the field. For imaging over areas larger than the
instantaneous field-of-view, it is convenient to use a footprint that can tile
the area to be imaged. Figure \ref{footprint} shows a footprint commonly used
for BETA observations.

\paragraph{Field rotation and the antenna roll axis} 

The roll axis is used to compensate for the field rotation that is normal for
altitude-azimuthally (alt-az) mounted telescopes, fixing the footprint's position angle
(PA) relative to the celestial coordinates.  The roll axis has rotation limits of
$\pm$180 degrees. When either of these limits is reached, observations must be
interrupted while the antenna is rolled back 360 degrees. With a requested $\text{PA}
= 0$, this limit occurs on the meridian for sources north of the zenith, and
on the lower meridian for more southerly sources.

The position angle of the footprint controls the orientation of the
instrument's polarization angle on the sky. The PAF elements receive
orthogonal dual linear polarizations (94 elements for each); these are
oriented at $\pm$45\degr\ to the vertical when $\text{PA} = 0$.

\paragraph{Beam separation and interleaving}

The choice of footprint pitch involves several considerations:
\begin{itemize}

\item Setting a large pitch samples a larger portion of the PAF field of view, but leaves
sensitivity depressions between the beams.

\item Setting a small pitch results in a more even sensitivity, but possibly loses sensitivity
because of correlation between beams.

\end{itemize}

The correlation between beams arises from a PAF element and its receiver noise
contributing to two or more beams. As the separation between two
electronically formed beams decreases, the number of PAF elements common to
both increases, and so does the correlation between their noise contributions.
We discuss this correlation in a later section.

To satisfy both of the constraints above, we use an interleaving technique,
whereby the synthesis observation is divided into two parts $A$ and $B$. A
comparatively wide pitch is chosen---typically equal to the FWHM at the
shortest wavelength in the observing band, and antenna pointing for $A$ and
$B$ is adjusted to place the beam maxima for $B$ on the sensitivity minima for
$A$. Figure \ref{footprint} illustrates this scheme. A broader discussion of
the interaction between interleaving and sensitivity over the field of view is
given by \cite{Bunton:2010}

Note that this scheme, with minimum beam spacing $\sim\text{FWHM}/\sqrt{2}$,
does not fully Nyquist-sample the sky. BETA's ability to reconstruct low
spatial-frequency brightness structures is already limited by its small number
of short baselines, so the wide beam spacing was chosen to maximise the field
size visible with only nine beams.

\paragraph{Calibration}

In general the antenna-specific complex gains of the telescope are expected to
be beam-dependent as each beam is composed of a unique combination of PAF
elements. Therefore each beam has its own bandpass response and amplitude
scale. With BETA, the practice has been to observe the flux-density standard
\PKSB\ for about five minutes with each beam, and to use these data to
calibrate each bandpass and set the flux-density scale.

Polarimetric observations with BETA have been calibrated for XY phase either
with an observation of the strongly polarized source 3C138, or with an
observation of the unpolarized \PKSB\ \citep{1984MNRAS.208..409K} but with the
roll axis of one antenna intentionally misaligned by 5\degr. The latter
technique leads to a known additional leakage between X and Y that can be used
to calculate the instrumental XY phase.

\section{CALIBRATION, IMAGING AND MOSAICKING}
\label{calibration}

Here we summarize the analysis steps taken for typical observational projects
with BETA. We refer the reader to descriptions of calibration, imaging and
mosaicking procedures used with BETA for continuum imaging
(\citeauthor{Hobbs:2016hi} \citeyear{Hobbs:2016hi},
\citeauthor{Heywood:2016di} \citeyear{Heywood:2016di}), spectral imaging
\citep{Serra:2015gz} and a search for spectral absorption over wide bands
\citep{2015MNRAS.453.1249A}. Analysis for these projects was performed using
several standard software packages: \textsc{miriad} \citep{Sault:1995ub},
\textsc{CASA} \citep{2007ASPC..376..127M}, \textsc{MeqTrees}
\citep{Noordam:2010jr}. Together, the results of these analyses provide
comparisons for the continuing commissioning of ASKAPsoft.

Although the ultimate intention is to reduce ASKAP data to wide-field images
in a single pass by gridding visibilities from all beams with AW-projection
\citep{Bhatnagar:2008gn}, BETA data have been processed beam by beam with
wide-field images produced in a final linear mosaic. The summarized procedure
follows.

\paragraph{Pre-processing} The calibration and target field measurement sets
are split to produce files of beam-specific data. Data are further split into
sub-bands according to specific requirements; for example the spectral
absorption processing separated the spectrum into 4-MHz or 5-MHz bands
corresponding to the beam-weight bands used in the digital beamformers (see
\textsection\ref{beamforming}). The data, still at full spectral resolution,
are checked and any with discrepant values are flagged.

\paragraph{Bandpass and flux-density calibration} Complex gains across the
band are determined from the \PKSB\ data, for each beam, using the flux-density
model of \citet{Reynolds:1994vd}. These gains are applied to the
target data, calibrating the bandpass and setting the flux-density scale. This
step also provides a good first estimate for the phase calibration across the
array. For continuum imaging, the data are now averaged to 1-MHz channels.

\paragraph{Develop a source model} Typical ASKAP fields are crowded,
containing tens of detectable sources per beam. Therefore we generate a model
of the field of each beam for further calibration to track gain variations
over the course of the observation. This model is derived from existing
records of the field, typically NVSS \citep{1998AJ....115.1693C} or SUMSS
\citep{Mauch:2003ed}, multiplied by the assumed BETA primary beam, or from an
initial image produced from the target dataset itself, or from a combination
of the two. As mentioned by \citet{Heywood:2016di}, the ASKAP roll axis holds
the primary beams fixed on the sky, and sources in side-lobes can be well
imaged and should be included in the field model used for calibration.

\paragraph{Self-calibration} All target data are calibrated (usually phase
only) using the field model. In some cases, the cycle of model generation and
calibration is repeated with a shorter calibration time interval.

\paragraph{Imaging and deconvolution} BETA operates with a single reference
direction for phase and delay tracking, common to all beams and set on the
boresight. Care is taken to either adjust the visibilities for each beam to
shift the phase centre to the beam pointing, or to centre the image on the
beam centre, not the phase centre. Standard imaging procedures are followed
using weighting schemes appropriate to the aims of the observation.

\paragraph{Mosaicking} Analytic models of the BETA beams are used to generate
weights in the linear mosaic. Some attempts have been made to use empirical
models derived from holographic beam measurements, but without any evidence of
improved results. We use a variation of the standard linear mosaicking
weighting scheme (linearly weighting each pixel by the inverse of the
variance---\citeauthor{1988A&A...202..316C}, \citeyear{1988A&A...202..316C}) to
account for the non-independence of image noise across beams. This
correlation, mentioned in \textsection\ref{observations}, arises from the
overlap of PAF element weighting functions for (especially adjacent) beams.
The amplifier noise from any single element will contribute to any beam
that has a non-zero weight for that element. In the presence of correlation
between beams and for the general case of a spectral image cube, the mosaic
noise is minimised by weighting as \citep{Serra:2015gz}

\begin{equation}
I_{\text{mosaic}}(l,m,\nu) = \frac{\mathbf{B}^T(l,m,\nu)\mathbf{C}^{-1}(\nu)\mathbf{I}(l,m,\nu)}{\mathbf{B}^T(l,m,\nu)\mathbf{C}^{-1}(\nu)\mathbf{B}(l,m,\nu)}
\end{equation}
where $\mathbf{I}$ and $\mathbf{B}$ are $N\times 1$ matricies representing the $N$ cubes and $N$
beams, and $\mathbf{C}$ is the $N\times N$ noise covariance matrix. \citet{Serra:2015gz} describe
using their continuum-subtracted data over frequency channels without H\textsc{i} emission to
calculate estimates of $\mathbf{C}$. For adjacent beams (separated by $\sim0.7 \times$~FWHM) they
report correlation coefficients in the range 0.13 to 0.2, and negligible correlation between more
distant pairs of beams.

\section{BETA PERFORMANCE}
\label{performance}

Here we summarise the results of many observations made with BETA, both measurements aimed at
characterising the telescope's performance and observations made with astronomical goals. 

\subsection{Sensitivity}
\label{results_spectrum}
\subsubsection{System Equivalent Flux Density}
\label{systemp}

\begin{figure}[h]
\begin{center}
\putfig{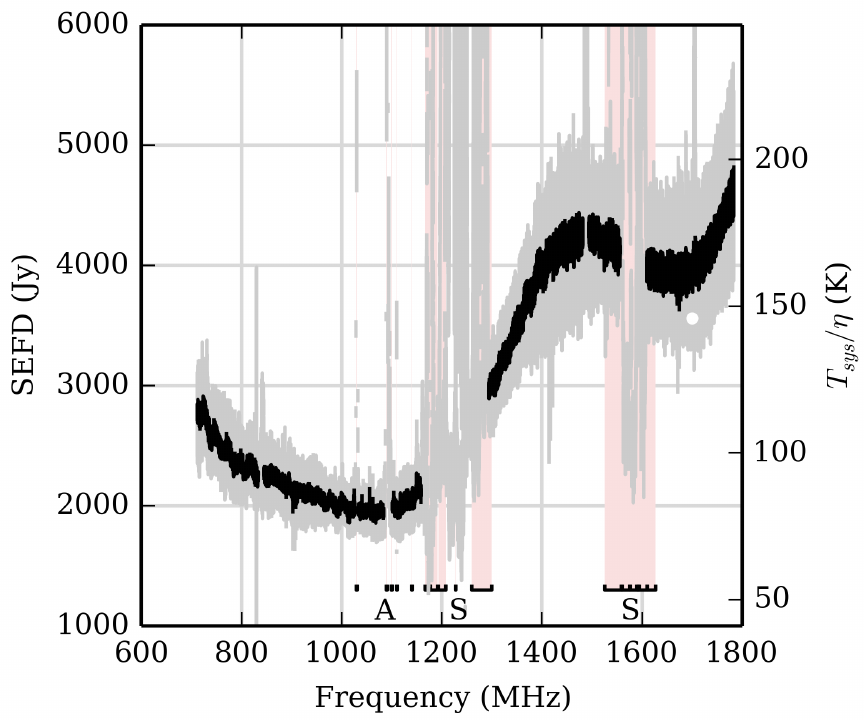}{\columnwidth}

\caption{The BETA spectrum of the System Equivalent Flux Density (SEFD)
computed as the standard deviation of real and imaginary components of the
visibility products, scaled by the measured amplitude of calibrator \PKSB\ and
the factor $\sqrt{2 \tau \Delta f}$, and decomposed into the antenna-specific
quanities. The right-hand scale gives the corresponding apparent system
temperature as $T_{\text{sys}}/\eta = \frac{A}{2k} \text{SEFD}$. The data shown here
are for a formed boresight beam, calculated for each 18.5\,kHz channel over a
980-s observation: the mean values over five antennas (AK09 was inoperable)
are shown in black, and the grey band indicates the ranges. The frequencies of
known radiofrequency interference are indicated by the bars below the plot,
labelled A (aircraft navigation) and S (communications and navigation
satellites).}

\label{sefd}
\end{center}
\end{figure}

Figure \ref{sefd} shows system noise in the portion of the radio spectrum
accessible to BETA (and ASKAP). The instantaneous bandwidth of the telescope
is 304\,MHz and this plot was generated from four separate observations. The
System Equivalent Flux Density (SEFD) is related to the equivalent temperature
$T_{\text{sys}}$ of the system noise by $\text{SEFD} = T_{\text{sys}} \frac{2k}{A\eta}$,
where $A$, $k$ and $\eta$ are the antenna area, Boltzmann's constant, and the
aperture efficiency. The SEFD was estimated from the variance of the real and
imaginary components of the visibility after being calibrated against the
flux-density calibrator \PKSB. Figure \ref{sefd} shows the results averaged
over five antennas, and can be compared with \citet[Figure 7]{Hotan:2014dv},
which shows an estimate of $T_{\text{sys}}/\eta$ determined on a single antenna. The
two measures are in broad agreement, differences being attributable to small
performance differences between PAFs and the different techniques used for
each determination. The high level of system noise above 1.2 GHz is a
consequence of impedance mismatch between chequerboard elements and their low
noise amplifiers \citep{Hotan:2014dv} and motivated the development of the
Mark II PAFs to be used on ASKAP. Early measurements of the Mark II PAFs
\citep{Chippendale:2015wyex} already show much reduced system noise across the
upper half of the band.

To obtain estimates of $T_{\text{sys}s}$ and aperture efficiency $\eta$
independently, we developed the new method described below. Normally, this
separate determination is accomplished with two calibrated temperature sources
to estimate $T_{\text{sys}}$, and an observation of a source of known flux
density to measure the effective area $A_{\text{e}} = \eta A$. The difficulty
of providing calibrated noise sources to the PAF receivers has led us to
develop a method using the known distribution of sky brightness temperature at
1.4~GHz \citep{Calabretta:2014fm,Alves:2015jr} and performing drift-scans---keeping antenna
pointing fixed and measuring visibilities as the Galactic Plane moves through
the beams. The system noise was estimated from the variance of the real and
imaginary components of the visibility products.  Discrete sources drifting
through the beam produced sinusoidal variations in the visibility amplitudes
that were easily recognised and removed before calculation of the variance. The set of variances for the whole array were decomposed into the antenna-specific quantities used in the linear fit described below.

\begin{figure}[h]
\begin{center}
\putfig{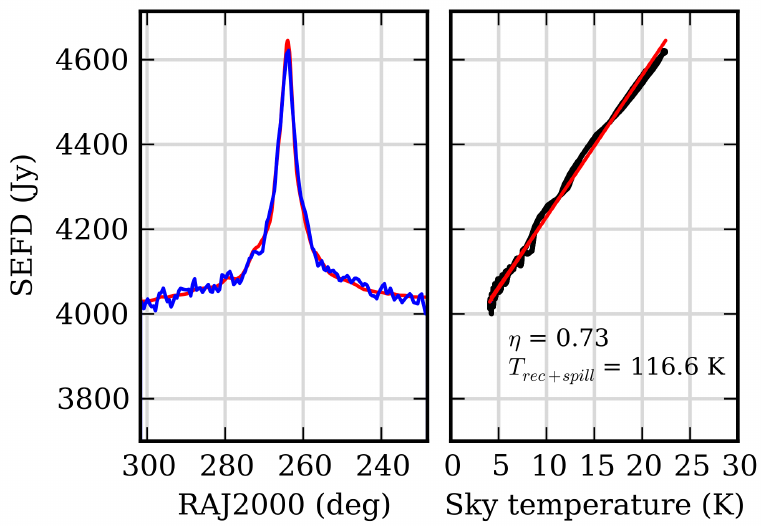}{\columnwidth}

\caption{Results of a drift-scan of the Galactic plane conducted on 2015 July
25 with antennas pointed at the southern meridian at zenith angle 5.4\degr;
these results are for a boresight beam on AK15 and were generated from an
8-MHz bandwidth centred at 1396~MHz. The left panel shows the variation of
SEFD during the scan (blue), and the flux-density equivalent of the fitted sky model (red); the
abscissa is labelled with the Right Ascension at beam centre. The right panel
shows the variation of SEFD with the 1.4~GHz sky brightness temperature as
determined from an all-sky Parkes continuum image \citep{Calabretta:2014fm}.
The red line shows the linear fit whose intercept and slope give
$\text{SEFD}(T_{\text{sky}} = 0)$ and $\frac{2k}{A\eta}$ respectively.}

\label{gpdrift}
\end{center}
\end{figure}

Figure \ref{gpdrift} displays the results for antenna AK15 for one such
measurement, and shows the expected linear relationship between system noise
and sky brightness temperature $T_{\text{sys}}$. The two parameters of the
linear fit yield $\eta$ and $T_{\text{sys}}^\prime = T_{\text{sys}} -
T_{\text{sky}}$. Measurements made over a range of zenith angles and with nine
formed beams per antenna have yielded 250 independent estimates of
$T_{\text{sys}}^\prime$ and $\eta$; their mean values are
$T_{\text{sys}}^\prime = 115\pm6 \text{K}$ and $\eta = 0.72\pm0.05$.


\subsubsection{Field of view}

\begin{figure}[h]
\begin{center}
\putfig{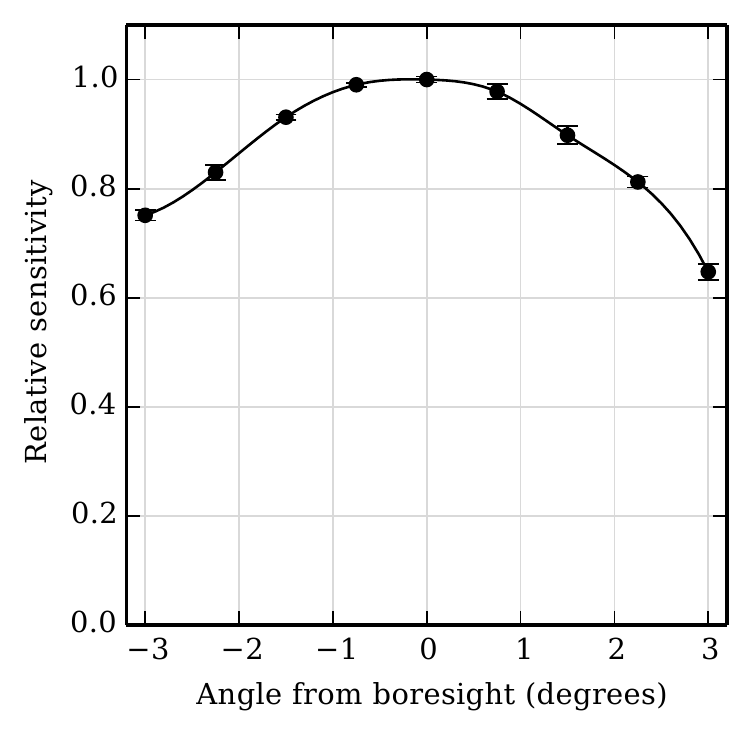}{\columnwidth}
\caption{
The sensitivity of beams arranged in a line across the PAF field of view,
relative to that of the boresight beam. The measurement was made from a single
observation of \PKSB\ with each beam in antennas AK03, AK08 and AK15. SEFD
values were computed for the band 960--980\,MHz from the visibilities on each
baseline in both X and Y polarizations; the mean of the quantity
$\text{SEFD}_0/\text{SEFD}_i$ for each beam $i$ is plotted. The error bars
indicate the variation across the three baselines.
}

\label{fov}
\end{center}
\end{figure}

To determine the variation of sensitivity across the Mark I PAF field of view,
we have measured the SEFD in each of a set of nine beams arranged linearly and
spanning the expected field of view. For the $i^{\text{th}}$ beam the relative
sensitivity is calculated as $\text{SEFD}_0/\text{SEFD}_i$ where
$\text{SEFD}_0$ is measured on the boresight beam.  The resulting profile
(Figure \ref{fov}) resembles that expected from electromagnetic modelling
\citep{Bunton:2010}.

\subsection{Beams}
\label{results_beams}

\subsubsection{Characteristics of maximum sensitivity beams}

\begin{figure}[h]
	\begin{center}
		\putfig{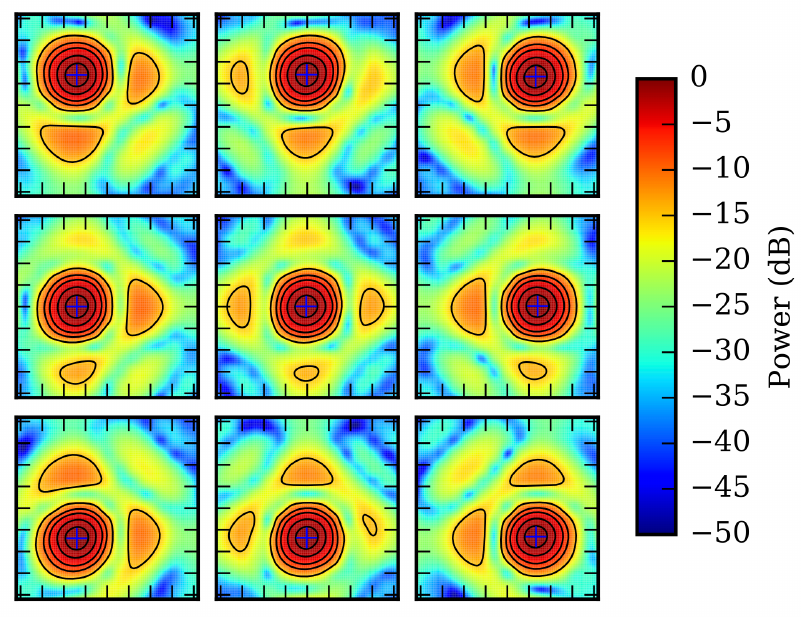}{\columnwidth}
		\caption{Holographic beam maps, each $8\degr \times 8\degr$, for a single 1-MHz channel with a frequency of 916\,MHz. The nine panels show 9 different beams, representing a square footprint with pitch 1.46 degrees. The two polarizations have been combined to form Stokes I. Contours represent 1, 3, 6, 9 and 15\,dB below the peak.}
		\label{SB1772_Mosaic}
	\end{center}
\end{figure}

\begin{figure}[h]
	\begin{center}
		\putfig{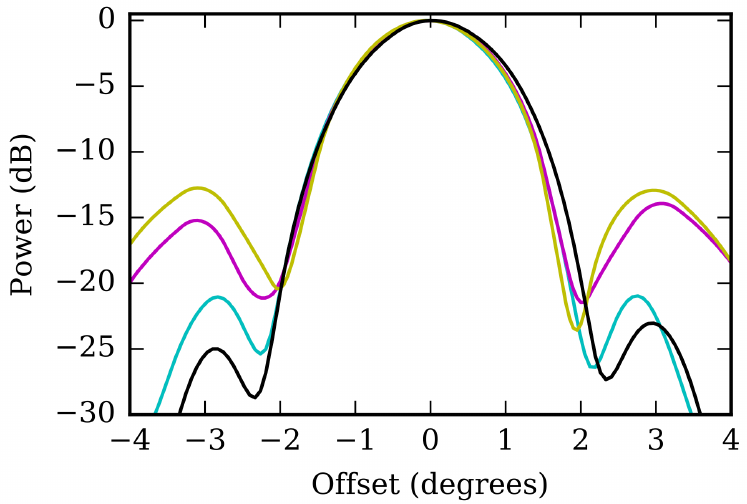}{\columnwidth}
		\caption{Slices through the map of the boresight beam shown in Figure \ref{SB1772_Mosaic}. The black and cyan traces are crossed diagonal slices through the centre, yellow is a horizontal slice and magenta is a vertical slice.}
		\label{SB1772_cuts}
	\end{center}
\end{figure}

Figure \ref{SB1772_Mosaic} shows holographic maps of the BETA beams.
Noteworthy features of the beams include strong four-fold symmetry
(particularly in the side-lobes) associated with the structure that supports
the PAF at the antenna focus, and elongation of the most offset beams due to
coma distortion. Because the antennas have a relatively small 12\,m aperture
and must support a PAF weighing approximately 300\,kg, the PAF itself and its
mechanical supporting structure create a significant aperture blockage that
clearly impacts the shape of the maximum sensitivity beams. On boresight, the
power in the side-lobes is roughly 15\,dB below the peak power level in the
primary beam in the worst case, but up to 10\,dB below this level in different
quadrants (see Figure \ref{SB1772_cuts}). The side-lobes become more prominent
for beams offset from centre.

The beams formed in the X and Y polarizations are noticeably elliptical with
major axes parallel to the plane of polarization. To quantify this, and to
explore the beam position and shape variation with wavelength, we have
analysed a set of central, boresight beams formed and measured on five of the
antennas over the 711.5--1015.5\,MHz range (the sixth was used as reference in
the holographic measurement). Although the holography naturally measures the
shape of an antenna's voltage response, we squared the measured amplitudes to
assess the pattern of response to incident power. There are 304 boresight
beams formed on each antenna: one for each coarse frequency channel. We
characterised each with an ellipse fitted to its half-power level, the
parameters of the ellipse giving the beam position relative to expectations
and the lengths of the beam major and minor axes.

We conducted this analysis on two sets of boresight beam, formed on 2015 May
08 and on 2015 May 12.  We found similar results for both, although the
holography observations of the first formed set suffered more from RFI.  In
the following paragraphs we give the results for the second set of beams.

\paragraph{Beam positions}

\begin{figure}[h]
	\begin{center}
		\putfig{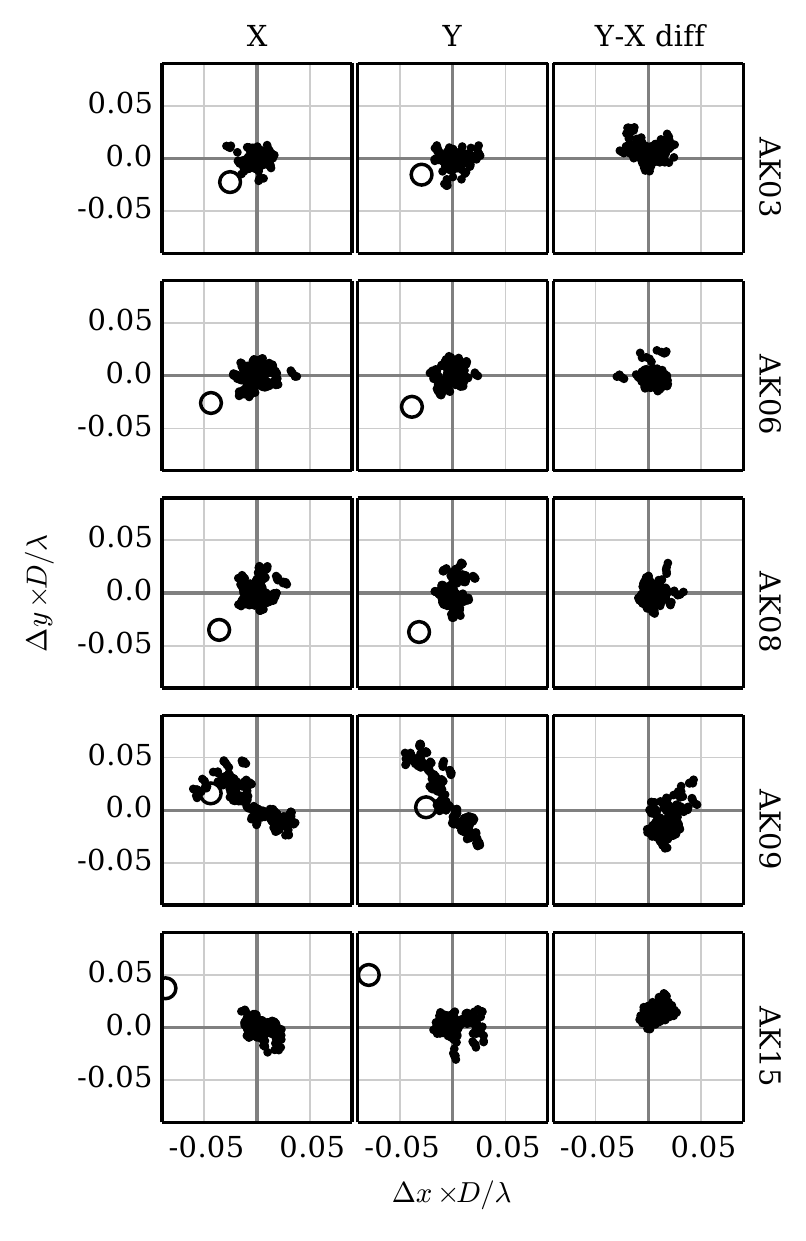}{\columnwidth}
		\caption{
		The left and centre columns show X and Y beam positions after
		subtraction of their median (marked by the open circle) across the
		711.5--1015.5\,MHz band, for five of the BETA antennas. The right column shows
		the positions of the Y relative to the X beam at each frequency.}
		
		\label{SB1803_beampositions}
	\end{center}
\end{figure}

Figure \ref{SB1803_beampositions} shows, for each antenna-polarization
combination, beam positions relative to their mean position, and scaled by
$\lambda/D$. Also shown is the position of each Y beam relative to the X beam
of the same frequency.  The median positions were subtracted before plotting
the $X$ and $Y$ values because they include contributions from antenna
pointing (irrelevant for this discussion) at the time of the holography
measurements.  From examination of the statistics of the measured beam
positions, we estimate the uncertainty ($1\sigma$) in position determination to be
approximately $0.0005 \lambda/D$, a value consistent with the strength of the
reference source and the System Equivalent Flux Density. For each antenna,
approximately 10 per cent of beam fits give discrepant values that we
attribute to the effect of intermittent RFI or other errors at the time of the
holography; these values fall outside the position ranges shown in Figure
\ref{SB1803_beampositions}. Amongst the remainder, significant and systematic
position variations are evident, with the behaviour being antenna dependent.

On most antennas, beams are clustered close to their median position with a
dispersion of about ten times the measurement uncertainty. For this set of
beams, antenna AK09 is an exception with a pronounced wavelength-dependent
position. In this display, any systematic position error common to all
spectral channels is hidden by the median subtraction. An indication of typical
systematic shifts is given in the third column of the figure that shows the
position difference between polarizations; over the five antennas, the median
position differences between X and Y beams range from 0.004 to 0.021
$\lambda/D$. At the centre of the observing band, these offsets are 0.4 -- 2.1
arcminutes, comparable to or larger than measurable antenna pointing errors
(see  \textsection{\ref{antennapointing}}).

\paragraph{Beam size and shape}

\begin{figure}[h]
	\begin{center}
		\putfig{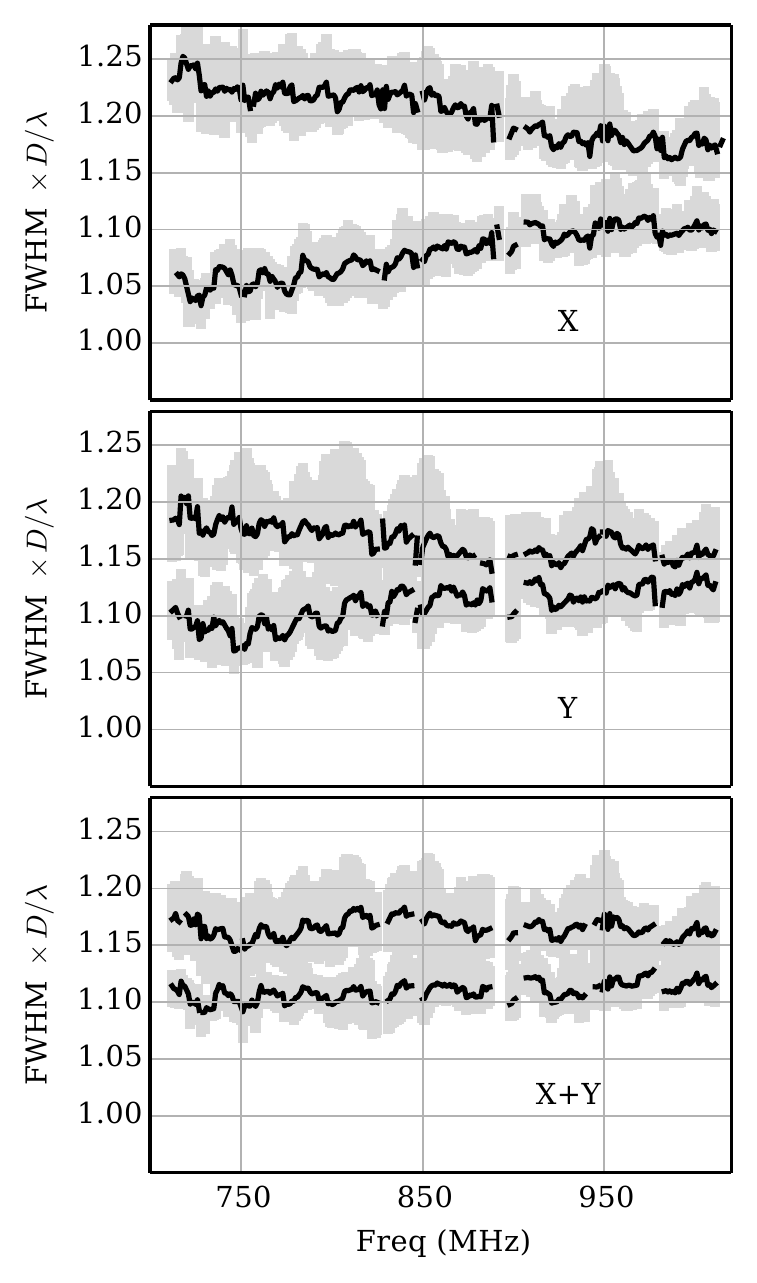}{\columnwidth}
		\caption{Beam shapes for X and Y polarizations and their sum (top, middle, bottom). The major and minor axes of best-fit ellipses at beam half-power level are given in units of $\lambda/D$. The heavy lines give the mean values over five of the six BETA antennas; the sixth, AK01 was the reference antenna in the holography measurements. The grey bands indicate the extreme values over the five antennas. Note the 25-MHz periodicity in beam width, which is referred to in the text.}
		\label{SB1803_beamshapes}
	\end{center}
\end{figure}

The beam widths and ellipticity, as determined from fitted half power levels,
are summarised in Figure \ref{SB1803_beamshapes}. The X and Y polarized beams
are elliptical with major axes parallel to the plane of polarization. The
ellipticity is wavelength-dependent. The total intensity beams (sum of X and Y
beams) are also mildly elliptical, but their widths are proportional to
wavelength. The figure shows a 5--10 per cent variation of beam widths amongst
the five measured antennas. Also evident is a 25-MHz periodicity in beam
width, which is the expected periodicity of a standing wave in the 6-metre
cavity between focus and vertex.

Together, the antenna and polarization dependencies of beam position and size produce a fractional
dispersion in antenna gains at the beam half-power points of approximately 10 per cent.

\paragraph{Modelling the beams}

The measured beams were formed using the maximum signal-to-noise ratio (maxSNR) method. This method
gives high weight to PAF elements in the brighter parts of the focused pattern, but it also
down-weights elements that have low gain or high noise, wherever they fall in the pattern.
Therefore, although the field distribution on the PAF may be well described by optical theory, the
actual beam formed will also depend upon the electrical characteristics of the individual elements
in the PAF. The variation in beam position and shape illustrated in Figures
\ref{SB1803_beampositions} and \ref{SB1803_beamshapes} is, in part at least, a consequence of the
variation in behaviour of the PAF elements.

Ellipticity of the polarized beams is expected. \cite{1968PIEE..115.1419M} have modelled the
fields in the focal plane of a circular paraboloidal reflector, uniformly illuminated by a linearly
polarized wave. They compute a field distribution that resembles the classic Airy pattern with a
central lobe and concentric lobes of alternating field direction, but which is elliptical with a
size and ellipticity dependent on the focal ratio $f/D$. The model as presented by
\citeauthor{1968PIEE..115.1419M} predicts the major and minor axes $(a,b)$, at half-power, of the
polarized beams to be $(1.11,1.02) \lambda/D$, and a circular total-intensity beam of width $1.06
\lambda/D$, all smaller than the measured dimensions shown in Figure \ref{SB1803_beamshapes}. We
have extended this model by modifying the illumination to include a circular central blockage and a
tapering function $T(\theta) = \cos(a\theta)$, where $\theta$ is the angular displacement from the
vertex as viewed from the focus, and the parameter $a$ determines the amount of tapering. We find
that the size of the total intensity beam (ignoring its small ellipticity) can be accommodated by the
model with a taper function $T$ that reduces the illumination at the edge of the antenna to about 44
per cent of its central value. Taper of that magnitude corresponds to an aperture efficiency of
$\eta = 0.70$, similar to the measured value given in \textsection{\ref{systemp}}.

We emphasize that this simple model for beam size and shape does not account
for asymetries in the optics, notably the tetrapod support structure for the
focal equipment. It assumes frequency-independent and circularly symmetric
radiation patterns for the PAF elements, almost certainly not realised. Nor
does it include effects of multi-path reflections. 

Analysis of the observed off-axis coma distortion is beyond the scope of this paper.

\subsubsection{Stability of formed beams}

Any variation in time of PAF element characteristics or signal path from PAF to beamformer has the
potential to also cause the beam shapes to change with time. During the early operation of BETA, it
became clear that beam performance was declining over the days following the determination of beam
weights.  The major factor leading to invalidation of a set of weights turned out to be random
delays of a few digital sampling intervals (1.3\,ns) being introduced with every power cycle of
the antenna pedestal hardware. This was sufficient to significantly degrade the quality of a formed
beam.

We adopted a two-fold solution to this problem (which will be fixed in firmware when the next
generation of ASKAP hardware is deployed). First, we determined a new set of maximum sensitivity
beams after every major power outage. Second, we took ACM measurements of a standard reference
field (a region centred on the South celestial pole) at regular intervals, which was used to make
a measurement of delays introduced per PAF element with respect to a reference epoch. If delays were
detected, they were compensated with an additional element-based correction to the digital
delay line.

\begin{table*}
\begin{center}
\caption[]{Fields with published images made with BETA.}
{\small
\begin{tabular}{|lllcccl|l|}
\hline
Field & \multicolumn{2}{c}{Position} & Band & Area & Type\textdagger & Reference \\
    & \multicolumn{2}{c}{(J2000) } & (MHz) & (Sq deg) & & \\
\hline
PSR J1107-5901 & $11^h07^m$&$-59\degr01$ & 711.5--1015.5 & 3 & CT & \cite{Hobbs:2016hi} \\
PKS B1740-517 & $17^h44^m$&$-51\degr30$ & 711.5--1015.5 & 20 & CS & \cite{2015MNRAS.453.1249A} \\
IC 1459 & $22^h58^m$&$-36\degr25$ & 1402 -– 1421 & 6 & S & \cite{Serra:2015gz} \\
Tucana & $\sim23^h$&$-55\degr$ & 711.5--1015.5 & 150 & C & \cite{Heywood:2016di} \\
\hline
\end{tabular}
} \\
{\footnotesize \textdagger C: Continuum; S: Spectral; T: Time series } \\
\label{images}
\end{center}

\end{table*}

\subsection{Polarimetry}
\label{results_polarimetry}

A series of observations were made in the 711.5--1015.5 MHz band to assess the
polarimetric performance of the BETA antennas. Two sources were chosen for the
tests: \PKSB, the standard southern flux-density calibrator known to have no
detectable linearly polarized emission, and 3C138, a well characterised and
strongly polarized source. These sources were observed on the optical axis of
the antennas; \PKSB\ was also observed off-axis, and again on-axis but through
eight different offset beams. For the tests of a beam on the optical axis,
observations were done both with the roll axis tracking (i.e. the normal BETA
mode) and with the roll axis locked (i.e. BETA antennas behaving as classical
alt-az antennas with parallactic angle rotation needing to be accounted for in
the polarimetric solutions). The two approaches gave consistent results.

For polarimetric work the phase difference between the X and Y polarization
channels must be known. Whereas some instruments measure this phase difference
using a noise injection system, such a system was not available with BETA (it
will be included in the full ASKAP system). For a significantly polarized
source, such as 3C138, XY phase differences can be determined as part of the
source observation. However, for an unpolarized source such as \PKSB, extra
steps need to be taken to make it possible to determine XY phase difference.
With BETA observations of \PKSB, the approach used was to intentionally
misalign the roll axis on one antenna (AK03) by 5\degr. This puts sufficient,
known, signal into the XY correlations of baselines with this antenna to be
able to determine the XY phase. This approach is somewhat akin to the
`cross-dipoles' approach used previously at the Westerbork telescope
\citep{Weiler:1973aa}. It was verified that the two approaches to determining XY
phase produce consistent results.
 
On-axis, BETA is polarimetrically pure. Polarization leakage between X and Y
channels was measured on both \PKSB\ and 3C138 and found, on most antennas, to
be small ($<$ 0.5 per cent) and frequency independent. The leakages were stable over
months. The only significant departure from zero leakage was a clear
positive-negative asymmetry between the real parts of the X and Y feeds in
some antennas, characteristic of a rotation of the whole PAF. These antennas
appear to have a small rotational misalignment about their roll axis; the
largest error observed was $\sim$2\degr in AK15. Once identified, such
misalignments are easily corrected.

For polarimetric measurements of 3C138, account must be taken of the Faraday
Rotation caused by the ionosphere. Ionospheric Faraday Rotation is a time-varying,
wavelength-dependent rotation of the plane of polarization (an
advantage of using an unpolarized test source such as \PKSB\ is that the
observation is not affected by the ionosphere). For 3C138 observations, the
ionFR package \citep{SotomayorBeltran:2013hk} was used with global GPS
ionospheric soundings to estimate the ionospheric rotation measure as a
function of time during the observation. After accounting for the ionospheric
rotation, the measured position angle and intrinsic rotation measure of
3C138's polarized emission was consistent with previously existing results.

Leakages measured with the source off-axis (placed at the 65\% level on the
axial primary beam) were similarly small, apart from $\sim$1\% leakage of the real
parts of X and Y signals near 1 GHz.

The main source of polarization inaccuracy in BETA is the {\em a priori}
unknown beam shape. This is particularly acute for the Stokes U parameter,
formed from the difference between XX and YY visibilities\footnote{Recall
that, in the standard observing mode, the elemental dipoles that make up the
BETA PAF are oriented at a PA of 45\degr; tracking the roll axis during an
observation keeps this angle constant.}. The analysis of a typical example of
formed beams described in \textsection{\ref{results_beams}} shows dispersion
in the shapes of X and Y beams, and in their relative positions. These both
impair the polarization performance, most severely for sources away from beam
centres. Little detailed analysis has been done on off-axis beams and the
implications of their properties on polarimetric performance.

More information on polarimetric characterisation of BETA is given by Sault
(\citeyear{Sault:aces2} and \citeyear{Sault:aces7}).

Because of the presence of polarized sources in the sky, the best imaging
results, even in total intensity (Stokes I), require attention to wide-field
polarimetric corrections. If all antennas can be treated as having
identical polarimetric responses that are independent of time, the
polarimetric correction can be relatively simple because the use of the roll
axis means that the polarimetric response does not vary throughout
the observation as parallactic angle changes. The instrumental polarimetric
response of a source is thus independent of time and baseline, and can be
corrected in the image domain after deconvolution. The correction would be
implemented by a Mueller matrix, the coefficients of which would be dependent
on a pixel's location relative to the pointing centre. This is analogous to
primary beam correction being performed in the image domain. However,
ionospheric Faraday rotation correction is a complication of such a scheme
because, in general, the ionospheric Faraday rotation will vary significantly
over the course of an observation. Correcting for ionospheric Faraday rotation
should be done before the imaging of the Stokes parameters, otherwise a time-
varying rotation between Stokes Q and U will be introduced. Approaches that
combine snapshot images are being investigated for implementation in ASKAPsoft.

\subsection{Image quality}
\label{results_image}

Many imaging experiments have been conducted with BETA. All have used beams formed with
the maxSNR method. A range of fields have been imaged, with various sizes, declinations and
complexity.  Several fields, such as the Apus field centred near (RA,Dec)$_{\text{J2000}}$ =
($16^h00^m$,$-78\degr30$), and the field containing calibration source \PKSB, have been used as test
fields and have been imaged many times during the commissioning period to allow assessment of
changes made to beamforming and calibration techniques. Images of four fields have been published
(see Table \ref{images}), and we draw upon the analysis of those images for our characterisation of
BETA's imaging performance below.

\subsubsection{Sensitivity}

The RMS brightness in spectral images with channel bandwidth of 18.5 kHz is consistent with the
measured SEFD at 1.4GHz of 4000 Jy (see Figure \ref{sefd}). However, the RMS brightness in continuum
exceeds that expected from the SEFD by a factor of about three. \cite{Heywood:2016di} attribute this
to the incomplete deconvolution that is inevitable with automated reduction pipelines in the
presence of the significant sidelobe confusion typical of BETA continuum images.

\subsubsection{Photometry}

Photometric performance has been assessed by comparing source catalogues constructed from BETA
images with catalogues from other instruments: the SUMSS catalogue \citep{Mauch:2003ed} for the
711.5--1015.5\,MHz image of the Tucana field, and the VLA NVSS catalogue \citep{1998AJ....115.1693C} for
the 1.4 GHz image of IC 1459. In both cases, systematic differences in the flux-density scales were
observed. Flux densities of sources in the IC 1459 image exceed those of the corresponding NVSS
sources by a factor of about 1.07. Apparent flux densities of sources in the Tucana image differ
from their counterparts in the SUMSS catalogue in a flux-density dependent way: the weakest sources
appear fainter in the BETA image, whereas the strongest sources appear brighter; the ratio varies
smoothly from $\sim0.84$ to $\sim1.04$ over the full flux-density range. \cite{Heywood:2016di}
discuss the possible causes of this discrepancy. By selecting the subset of sources that are
unresolved in the SUMSS catalogue, they exclude bias caused by the different PSFs of the two
instruments. Both \cite{Serra:2015gz} and \cite{Heywood:2016di} suggest that limitations in our
knowledge of beam shape can contribute to the observed bias.

We note that flux-scale differences amongst catalogues constructed with different instruments or
techniques are not unsual---see \cite{Allison:2014ht} for an example similar to the BETA-SUMSS
comparison. In general, precise photometry with radio interferometers is difficult. There are many
contributing factors, including statistical effects (eg. the Eddington bias), differences in spatial
frequency coverage, and the complex interactions between deconvolution and self-calibration that are
invariably instrument-specific. There is no evidence for the observed flux-scale discrepancies
indicate a fundamental limitation of ASKAP. In future the increase in numbers of antennas and
improved knowledge and control of primary beam shape are likely to improve photometric performance,
but ASKAP will remain subject to all the factors common to other radio telescopes.

\begin{figure*}
	\begin{center}
		\putfig{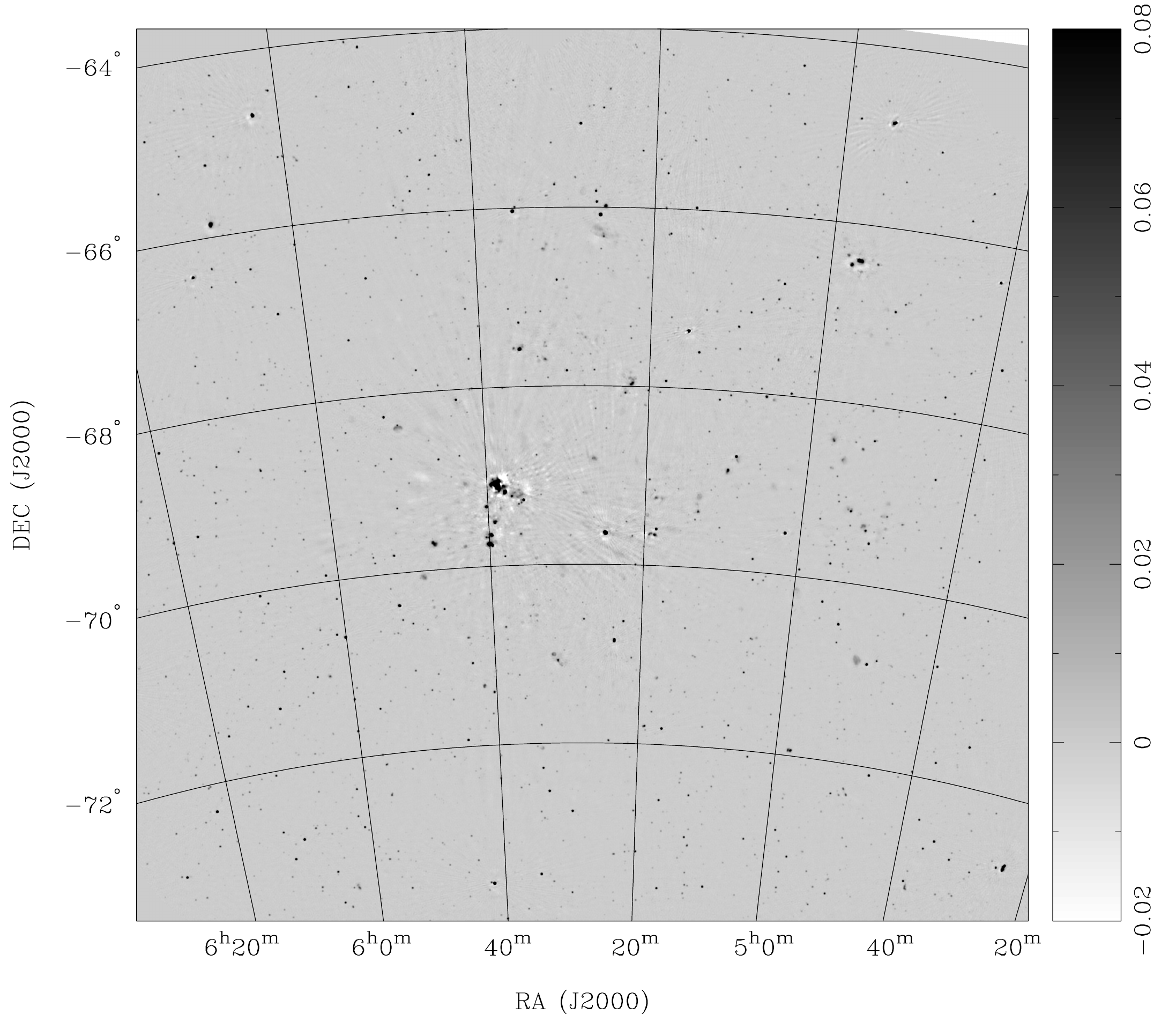}{\textwidth}
		\caption{Image of the Large Magellanic Cloud in the 711.5--1015.5\,MHz band. The 11-hour observation was
		made with eight pointings of a square 9-beam footprint: a pair of $2 \times 2$ grids, offset
		from each other to achieve the interleaving scheme illustrated in Figure \ref{footprint}.
		The eight pointings were observed cyclically, for five minutes in each cycle giving an
		integration time of 82 minutes on each pointing. The deconvolved image was restored with a 
		$60\arcsec\times60\arcsec$ beam, and the brightness scale on the right is in units of Jy/beam.
		Note the image artifacts associated with the bright and extended HII region 30 Doradus near
		(RA, Dec) = ($05^h38^m$,$-69\degr06\arcmin$).  This image was produced with an automated
		ASKAPsoft pipeline.}

		\label{lmcimage}
	\end{center}
\end{figure*}

\subsubsection{Astrometry}

\cite{Heywood:2016di} analysed BETA's astrometric precision and accuracy by comparing apparent
source positions in the Tucana image with those of the SUMSS catalogue. They separated the
differences into a random statistical component and a systematic shift, identified as the mean
offset in RA and Dec for each of the three epochs of the Tucana observation. Once the systematic
offsets are subtracted, the distribution of residual errors is similar for all three epochs and
broadly consistent with the size of the synthesised beam ($70\arcsec\times60\arcsec$) and the
signal-to-noise ratios of the sources in the BETA image. \cite{Heywood:2016di} report the
$1\sigma$ uncertainties in (RA,Dec) to be (3.6,4.5) arcseconds, averaged over the three epochs.

The systematic offsets were different for each epoch; the largest had a magnitude of 12.7
arcseconds. Astrometric accuracy of a radio interferometer is usually achieved through calibration
of interferometer phases with observations of a source with well-known position and that is close to
the observed field. BETA's imaging observations were calibrated with a single observation of
\PKSB\ before or after the twelve-hour synthesis. The systematic position errors observed can be
understood in terms of small imperfections in the telescope's delay (antenna position) model and the
transfer of phase reference from a calibrator 25 degrees distant, and in likely electronic drifts in
the PAF amplifiers.

In the future, a model of the sky, a `Global Sky Model' \citep{Cornwell:2011}, will be developed and
used to calibrate antenna gains. Inclusion of astrometric calibrators in this model, with a sky
density sufficient for at least one to appear in every ASKAP field, will allow all observations to
be tied to an astrometric reference frame.

\subsubsection{Response to complex source structure} Images of complex fields
have been made, such as the Large Magellanic Cloud (LMC) shown in Figure
\ref{lmcimage}, the bright radio galaxy NGC 1316 in Fornax, and the region of
the Galactic plane close to PSR J1107-5901 \citep{Hobbs:2016hi}.  Figure
\ref{lmcimage} shows an image of the LMC made in the 711.5--1015.5\,MHz band
with visibilities from the 37-metre baseline excluded. This image was produced
using ASKAPsoft, following a series of steps similar to those listed in
\textsection{\ref{calibration}}. In this case, data were preconditioned using
a Wiener filter (see \citeauthor{Cornwell:2011} \citeyear{Cornwell:2011} for
an explanation of methods used in ASKAPsoft) to effect a traditional robust
weighting with parameter $r = -0.5$ \citep{Briggs:1995}. The image was
deconvolved using a multiscale procedure: ASKAPsoft's `BasisfunctionMFS'
algorithm with scales of 0, 3, 10 and 30 pixels. The complex extended radio
emission associated with the star forming region 30 Doradus (near
$05^h38^m$,$-69\degr06\arcmin$) is not well represented in this image: BETA's
shortest baselines (37, 144, 176 metres) give inadequate sampling of the inner
part of the $(u,v)$ plane for reconstructing structures of this scale, as
discussed by \cite{Serra:2015gz}.

\subsubsection{Beam variations}
All imaging observations with BETA have been made with maxSNR beams, whose typical characteristics
we present in \textsection{\ref{results_beams}}. Continuum images made with BETA do not achieve the
sensitivity expected from thermal noise considerations. \cite{Heywood:2016di} attribute this
sensitivity loss to calibration biases and incomplete deconvolution that are difficult to avoid with
a small array like BETA. Another contributor to sensitivity loss is the dispersion in beam position
and size reported in \textsection{\ref{results_beams}}. Ideally, the gain of the $i^{\text{th}}$ antenna (for
a given polarization) could be written as functions of time $t$, frequency $\nu$ and direction
$(\theta,\phi)$:
\[
g_i(\nu,t,\theta,\phi) = A(\nu,\theta,\phi) b_i(\nu) f_i(t)
\]
where the bandpass $b_i$ and the normalised antenna reception pattern $A$ are both time-invariant
and $A$ is the same for all antennas in the array with its angular scaling inversely proportional to
frequency. Accounting for variation in $A$, with time, polarization or antenna is difficult and
expensive in computation time, and has not been implemented in any of the automated BETA processing
software. Uncorrected, variation of $A$ will decrease the dynamic range of the image.

\section{SUMMARY AND CONCLUSIONS}
\label{summary}

BETA has been used to develop methods for operating, calibrating and making
astronomical observations with a synthesis telescope equipped with phased
array feeds. The focus of the work has been learning how to electronically form
beams, how to measure and arrange beams, and how to modify the familiar steps
of calibration and synthesis imaging to make the best use of formed beams.

A range of astronomical imaging and spectral detection experiments have been conducted in support of
commissioning, and to take advantage of BETA's wide field of view and its access to the relatively
clean part of the radio spectrum from 0.7 to 1.2 GHz. The scientific results from these are
summarised in \textsection{\ref{firstscience}}.

We began in the introduction (\textsection{\ref{sec:intro}}) with some
questions about the future success of ASKAP for which we sought answers through
experiment with BETA.  Several of the questions have been answered positively;
for others the BETA experience has provided partial answers with good
indications of the remaining problems to be solved.  We conclude by listing
three critical challenges to be met in the operation of ASKAP:
\begin{description}
\item[Reliability] Large-area surveys conducted over
months or years demand a high level of stability and reliability in the
telescope.  For much of its operation, BETA's reliability was sufficient to
yield uniform results for modest-sized surveys.  ASKAP will have many more
components (antennas, beamformers) and modes of operation and its surveys will
be much larger.  These will all place much higher demands on reliability than
were necessary for BETA.
\item[Beam shape] We have described the maxSNR beam
forming methods used with BETA and assessed metrics of the resulting beams.
We continue to develop alternate methods that give control over beam shape;
in spite of promising progress, more work is required to prove these methods
and to make them operational.  Successfully constraining beam shape is important
for success of several of ASKAP's SSPs, particularly so for wide-field
polarimetry.

\item[Gain stabilisation] BETA had no reference noise signal for measuring
receiver (PAF element) gains.  However, for the observations made with BETA,
gains were stable enough to allow synthesis imaging with a single measurement,
per beam, of a standard source for bandpass calibration, and with minor
corrections to antenna gain variations during the synthesis using self-calibration
with source models of the field. Daily (or more frequent) observations for
calibrating each of ASKAP's 36 beams will significantly degrade observing
efficiency, so methods for using the on-dish-calibration system for measuring
PAF element gains will need to be developed.
\end{description}

\begin{acknowledgements}

The Australian SKA Pathfinder is part of the Australia Telescope National
Facility which is funded by the Commonwealth of Australia for operation as a
National Facility managed by CSIRO. This scientific work uses data obtained
from the Murchison Radio-astronomy Observatory (MRO), which is jointly funded
by the Commonwealth Government of Australia and State Government of Western
Australia. The MRO is managed by the CSIRO, who also provide operational
support to ASKAP. We acknowledge the Wajarri Yamatji people as the traditional
owners of the Observatory site.

Parts of this research were conducted by the Australian Research Council
Centre of Excellence for All-sky Astrophysics (CAASTRO), through project
number CE110001020.

This work was supported by resources provided by the Pawsey Supercomputing
  Centre with funding from the Australian Government and the Government of
  Western Australia.
\end{acknowledgements}

\raggedright
\bibliographystyle{hapj}
\bibliography{full}

\end{document}